\documentclass[12pt]{article}
\usepackage{amsmath,amsfonts,graphicx}
\newcommand{\Alfven}{$\rm Alfv\acute{e}n$}
\newcommand{\D}{\partial}
\newcommand{\DD}{\frac}
\newcommand{\TAW}{\tiny{\rm TAW}}
\newcommand{\mm }{\mathrm}
\newcommand{\Bp }{B_\mathrm{p}}
\newcommand{\Bt }{B_\mathrm{T}}

\newcommand{\beq}{\begin{equation}}
\newcommand{\eeq}{\end{equation}}
\newcommand{\ben}{\begin{enumerate}}
\newcommand{\een}{\end{enumerate}}

\newcommand{\ber}{\begin{array}}
\newcommand{\eer}{\end{array}}

\newcommand{\cd}{\!\cdot\!}
\newcommand{\Msun}{{\,{\cal M}_{\odot}}}

\newcommand{\Mdot}{{\,\dot{\cal M}}}
\newcommand{\degr}{{}^{\circ}}
\def\Title#1{\begin{center} {\Large {\bf #1} } \end{center}}

\begin{document}

\Title{A model for the jet-disk connection in BH accreting systems}

\bigskip\bigskip

\begin{raggedright}  
 
{\it A. Hujeirat${}^1$, M. Livio${}^2$, M. Camenzind${}^3$ and A. Burkert${}^1$  \\
${}^1$Max-Planck-Institut f\"ur Astronomie, 69117 Heidelberg, Germany  \\
${}^{2}$SSTI, 3700 San Martin Drive, Baltimore, MD 21218, USA, \\
${}^{3}$Landessternwarte-Koenigstuhl, 69117 Heidelberg, Germany}
\bigskip\bigskip
\end{raggedright}

\section{Abstract}
  The powerful and highly collimated jets observed in active galactic nuclei and  $\mu-$quasars
  are likely to be connected to the accretion phenomenon via disks.
  Based on theoretical arguments and quasi-stationary radiative MHD calculations, 
  a model for an accretion-powered jet is presented. It is argued that accretion disks
  around black holes consist of 1) a cold, Keplerian-rotating and weakly magnetized medium
  in the outer part, 2) a highly advective and turbulent-free plasma inside 
  $r_\mm{tr} = 10-20$ Schwarzschild radii, where magnetic fields are predominantly
  of large scale topology and in excess of thermal equipartition, and 3) an ion-dominated
  torus in the vicinity of the hole, where magnetic fields undergo a topological change into a
  monopole like-configuration. The action of magnetic fields interior to $r_\mm{tr}$ is
  to initiate torsional {\Alfven} waves that extract angular momentum from the disk-plasma 
   and deposit it into the transition layer between the disk and the overlying corona,
  where the plasma is  dissipative and tenuous.
  A significant fraction of the shear-generated toroidal magnetic field reconnects
  in the transition layer, thereby heating the plasma up to the virial-temperature
  and forming a super-Keplerian rotating, and hence centrifugally accelerated outflow.\\
  The strong magnetic field in the transition layer forces the electrons to 
  cool rapidly which, in combination with the fast outward-oriented motion, yields a
  two-temperature ion-dominated outflow. The toroidal magnetic field 
  in the transition layer is in thermal equipartition with the ions, whereas the poloidal
  component is in equipartition with the electrons. Such a strong toroidal magnetic field
  is essential for increasing the jet-disk luminosity in the radio regime.
  These gravitationally unbound outflows  serve as seeds, possibly, 
  for all the powerful electron-proton jets observed in accreting systems containing black holes.
\section{Introduction}
 Based on observational data, most of the systems containing jets are considered 
to be accreting systems, most likely with an accretion disk surrounding the central object. 
The most powerful and highly collimated jets are found to emanate from accreting systems
 containing, probably, black holes, with a maximum power attainable if the central
 objects are  Kerr black holes (-BHs) rotating at their maximum rate \cite{Blandford4}. Beside the numerous extragalactic radio sources
 with high gamma-factors ($\gamma \ge 3$), the underlying engines powering the jets in microquasars
 such as  GRS 1915 + 105 and GRO 1655 - 40 are believed to be  
spinning BHs \cite{Mirabel01}. 
 
Although there has been a significant progress in recent years
toward understanding the morphology, propagation and shock diagnostics of jet-plasmas,
no consensus has been reached yet about their basic driving mechanisms.  
Nevertheless, there are at least three ingredients that appear to be necessary 
for initiating jets that gained  theoretical and observational supports: 
1) large scale magnetic fields (-MFs), 2) an accretion disk and 3) a central object which
 dominates the disk dynamics gravitationally.
  Magnetic fields in particular are considered to play a major role in 
 the evolution of jets. The fact that many jets have comparable radio and bolometric
 luminosities  hints to a significant fraction of the magnetic energy in their total power.\\
   In the first place MFs extract rotational energy from the underlying disk
   and deposit it into the plasma at higher latitudes, thereby initiating motions that are
 principally outward-oriented.  \\
 At relatively large radii MFs are pre-dominantly toroidal, and their main 
 action is to force the rotating particles to collimate around the corresponding axis of
  symmetry.

The central object must be relatively massive compared to the disk, so that 
  its gravitational well is sufficiently deep for initiating motions with speeds 
  comparable to the escape velocity near its surface.
   
In this paper we present a model for initiating jets from accretion disks around BHs.
The model is a modified version of the truncated disks advective tori model (TDAT; \cite{HujCb},
in which a special attention is given to the role of MFs.  

Our approach relies on theoretical arguments  
       supported by the radiative two-temperature MHD numerical calculations. The model is based on
      the following assumptions: 1) The Balbus-Hawley instability acts as dynamo that amplifies the unordered MFs up
      to thermal equipartition, 2) The Parker- and BH-instabilities in combination with reconnection
      close a dynamo cycle,  through which 
      a large scale  MF is generated, in the manner that Tout \& Pringle (1992) suggested.
      Once the large scale magnetic field grows beyond thermal equipartition, the 
      generation and dissipation of turbulence will be suppressed.  
      3) A significant fraction of the toroidal magnetic field (-TMF) lines re-connects in the
      transition layer (-TL): a process which 
        is treated      
       by adopting a turbulent diffusivity $\eta_\mm{mag}$, 4) Turbulent dissipation preferentially  heats the 
      heavy particles rather than the electrons, so that the two-temperature description for the plasma 
      can be applied\cite{Rees}.
      Fig. 2 shows the main phases in the evolution of a accretion-induced jet.\\ 
The paper runs as follows. In Sec. 3 we review several model for jet initiation, and outline
the necessity for a new model. The structure of magnetized disk is discussed in Sec. 4. 
In sec. 5 the governing equations to be solved and the method of solution are described, while the corresponding results
are presented in Sec. 6, and end up with the summary of the results in Sec. 7.
\begin{figure}[htb]
\begin{center}
{\hspace*{-0.5cm}
\includegraphics*[width=6.5cm]{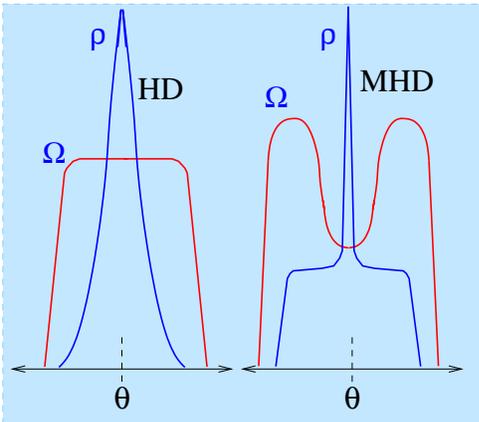}
}
\end{center}
{\vspace*{-0.4cm}}
\caption [ ] { A schematic distribution of the rotational velocity $\Omega$ and density $\rho$
              across the disk with magnetic fields (MHD) and without (HD).
   The action of the
  equipartition MFs is to transport angular momentum from the disk
   to higher latitudes, where the particles 
  are centrifugal-accelerated outwards.
  } 
\end{figure}
\section{The necessity for a new model}
Several models have been suggested for initiating jets\cite{Ghosh}, \cite{Blandford2},
 \cite{Pudritz}, \cite{Uchida}, \cite{Shu88}, \cite{Shu94}, \cite{Koenigl}, \cite{Heyvearts},
\cite{Camenzind}, \cite{Camenzind2}, \cite{Lovelace}, \cite{HujCL}.
 In the following we outline several  properties
of a few of these models:
\ben
\item The Blandford\& Payne  model (1982, henceforth BP82) relies on magnetic
 extraction of angular momentum and rotational energy from an underlying cold and
 Keplerian disk, i.e., from a standard disk (-SSD, \cite{Shakura}).
 Extraction is most efficient if the poloidal magnetic field is of a large scale
 topology and appropriately aligned 
 to the disk-normal (i.e.,  the angle from disk-normal  $\theta_\mm{B}$ must be larger than
  $30\degr$). The second important result is that, at large radii, collimation 
 is achieved through the action of the shear-generated toroidal MF. 
 The difficulties of this model are:
\ben
  \item The disk is infinitely thin. This implies that Lorentz forces exerted on the disk-plasma
    would force the inflow to rotate sub-Keplerian, making it difficult for the plasma to be
flung out by purely centrifugal means\cite{Shu91}. Furthermore, taking into account
that toroidal magnetic field component $\Bt (z=0) = 0$ and adopting  the density profile $\rho \sim r^{-3/2}$ as BP82 suggested, any dynamo cycle
will not succeed to amplify $\Bp$ at $z=0$ to values beyond thermal equipartition. One possible 
way to obtain\footnote{$\beta = P_\mm{mag}/P_\mm{gas},$ where
  $P_\mm{mag} = (B^2_\mm{P}+B^2_\mm{T})/4\pi,$ see Eq. 8 for further details.}
$\beta >1$
 is through a collapse of the central part of the disk while freezing its 
 magnetic flux.  In this 
case\footnote{$\Phi = 2 \pi r^2 B_\mathrm{p} \approx 2 \pi r^2 B_\mathrm{\theta} = const.$}
  $\Bp \sim r^{-2},$ 
 which implies that if $\Bp$ does not change topology, accretion will be terminated 
 and the whole inflow will turn into outflow with unacceptably large mass load.

\item It has been argued that even if the poloidal MFs intersects the disk with
      $\theta_\mm{B} > 30\degr$, thermal assistance is still required for the
      disk-plasma to overcome the potential difference parallel to the MF-lines.
\item There is no special treatment to the flow in the deep gravitational well of the central object,
      which is necessary for increasing the total energy per gram of the inflow, and
      convert it into outflows with high speeds.
\item  The model relies on ideal MHD treatment to a cold Keplerian disk. 
       However, since $\beta \ge 1$, MFs will likely suppress the generation 
       and dissipation of turbulence. Hence, the disk is likely to be thermal unstable.
       If the accretion rate is high, the disk becomes optically thick to
       Synchrotron/cyclotron radiation; gyrating electrons emit radiation on the
       cyclotron frequency, that in turn repeatedly upscattered by the hot electrons, 
       establishing thereby a thermal equilibrium. Since this occurs on a time scale 
       comparable or even 
       shorter than the dynamical one, and in the absence of other efficient heating sources,
       the disk undergoes a runaway cooling: the temperature decreases and reach 
       the lower limit $T_\mm{min} = h\nu_\mm{peak}/k$, where $ \nu_\mm{peak}$ is the peak
       cyclotron frequency.  This thermally induced  collapse of the inner disk 
       gains acceleration if the accretion rate is low. Here the low frequency photons
       emitted by the gyrating electrons escape from the system without
       being absorbed or scattered, and so no lower limit for the temperature can be constructed.  
\een
\item The X-wind model (Shu et al. 1994). Here the magnetically driven-wind emanate, neither
      from the disk nor from the central object, but from 
     the region around the co-rotation radius, $r_\mm{cor}$, where the effective gravity
     vanishes.  The model relies primarily on the poloidal MFs of the central object which is surrounded 
     by an accretion disk, essentially without large scale magnetic fields. This approach relaxes
     the winding up problem of the poloidal MFs, which has been encountered in the 
     Ghosh \& Lamb model (1979). Moreover, as outflows here originate from the region close
     to the surface of the star, their propagation speed is estimated to be of the order of the
     escape velocity. \\
     When extending this model to accretion flows around BHs, the following problems arise:
\ben
\item It is not applicable to accretion flows onto Schwarzschild BHs, as these objects 
      do not possess dynamically stable poloidal MFs. If such holes are viewed as extreme
      stars, then the X-wind model rule out indefinitely the possibility of jet initiation from
       weakly magnetized slowly rotating objects.
\item Balbus \& Hawley instability amplifies weak MFs up to approximately thermal equipartition
      on the dynamical time scale. In the innermost region of the disk, MFs of the central 
      object are beyond super-equipartition with respect to the thermal energy of the 
      disk-plasma; generation and dissipation of turbulence will be 
      suppressed,  the plasma in the X-region (see Fig. 2b in \cite{Shu94} becomes cold,
      and the radial accretion will be  terminated. 
\item Almost all young stellar objects have been observed to rotate far below the beak-up
      velocity.
      Although the FU Orinois objects accrete at relatively high rates, they are considered
      to be slow rotators. Fast rotators, however, show excessive magnetic activities, which is 
      in line with the Parker instability{\cite{Simon}}.
      In the X-wind model however, MFs in the X-region are nearly in equipartition with the potential
      energy of the flow. Such strong magnetic tubes are likely to be gravitationally unbound, 
      cannot be anchored deep in the star and therefore they float upwards to the surface on the
      dynamical time scale. This occurs if they are generated in the convection zone or in the overshoot
      layer between the convection zone and the underlying rigidly rotating core, similar to the solar
      dynamo\cite{Spiegel}. 
\een   
\item Advection-dominated inflow outflow solutions -ADIOS (Blandford \& Begelman 1999), \\
      ADIOS are special case of ADAF solutions (advected dominated accretion flows) in which
      the accretion rate is  allowed to adopt the self-similar profile 
      $\Mdot \sim r^\mm{p}$, where $ 0\le p \le 1$.
      Thus, the accretion rate decreases inwards, thereby giving rise to a substantial
      outflow with large mass load. ADIOS model is another step closer than ADAF toward 
      explaining the low luminosity  AGNs. Indeed, there are several numerical calculations
      that confirm the inward decrease of the accretion rate (e.g., \cite{Stone}, \cite{Hawley}). 
     However,
      these calculations did not rule out the possibility that the outflows obtained might be
      large scale circulation.      
      On the other hand, in addition to the difficulties associated with ADAF solutions,
      ADIOS model:
\begin{figure*}
\begin{center}
{\hspace*{-2.2cm}
\includegraphics*[height=12.0cm,width=18.0cm, angle=0]{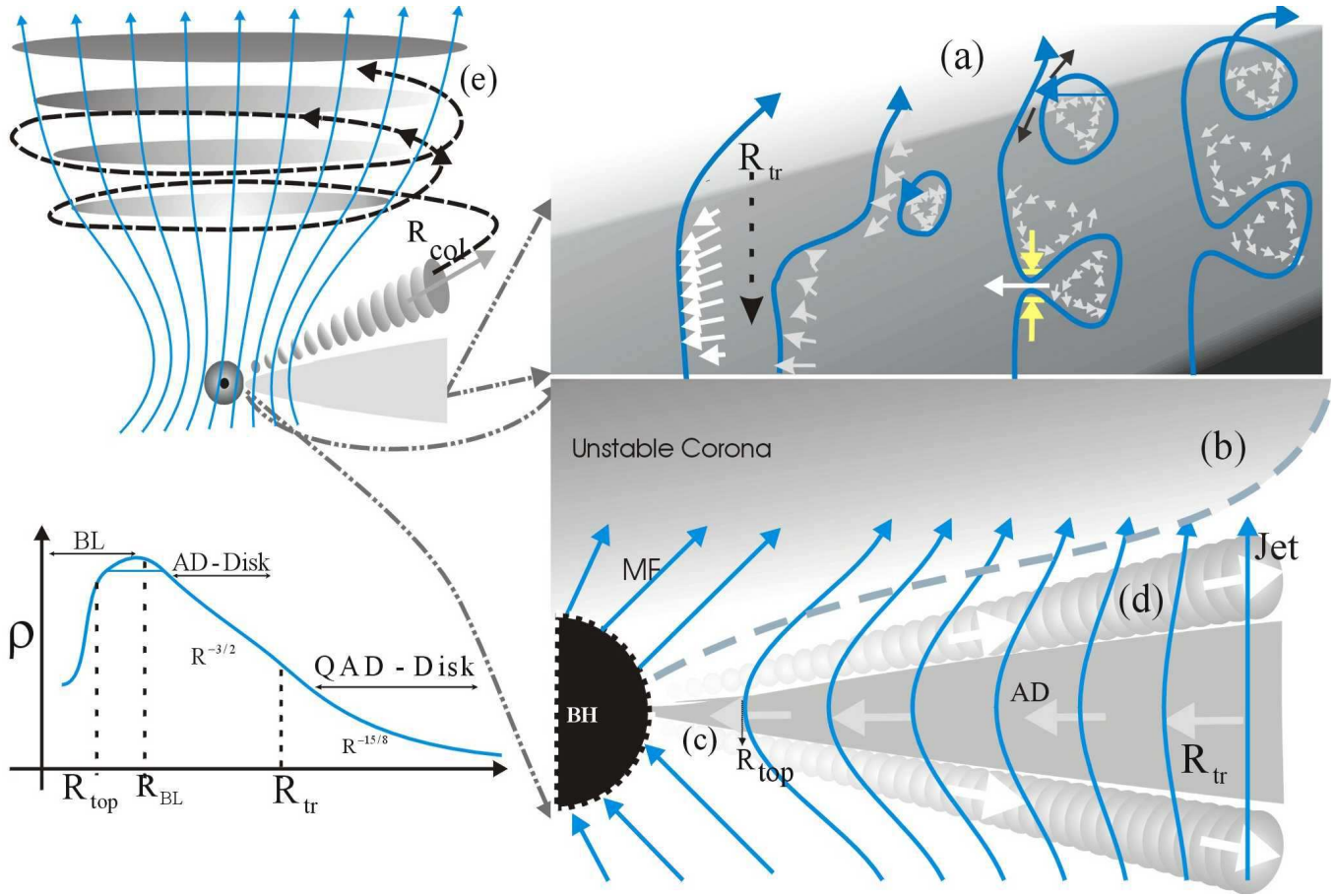}
}
\end{center}
{\vspace*{-0.4cm}}
\caption [ ] { A schematic illustration of the jet-disk connection.
 In the outer region of the disk MFs are weak and advected-around MFs by the fluid motions.
 BH- and Parker instability in combination with reconnection and inwards accretion
 amplified the MFs up to thermal equipartition. The MFs become of large scale topology
 and trun the disk at $r_\mm{tr}$ into advection-dominated (part a). Interior to $r_\mm{tr}$ (b), 
 turbulent-generation
 and dissipation are suppressed, and torsional {\Alfven} waves become the dominant angular momentum transporter (c).
 torsional {\Alfven} waves  carry angular momentum from the disk and deposit it into the thin layer above,
  where ions rotate super-Keplerian, virial-heated by reconnection of the TMF-lines, 
  and start to centrifugal-accelerate outwards (part d), 
  The ions in the TL are associated with internal, rotational and TMF-energy
  to start collimation at $r_\mm{col}$, where the magnetic-diffusivity becomes
  vanishingly small (part e). For clarity,   a schematic description of
   the density profile versus radius of the model is shown.  
  } 
\end{figure*}
 \ben
\item  relies on one dimensional treatment to an intrinsically multi-dimensional phenomenon.
       The model does not provide answers to numerous questions related to jet-morphologies, such as
       the origin of the high speed of jet-propagation, the role of the central object, the
       magnetic field topology appropriate for initiating jets and etc. 
\item  It largely overestimates the mass load of jets. Basically, it predicts that almost all
       accreted matter through the disk at large radii will re-appear in the jet. 
       When applied to the galactic center, 
       an accretion rate that is about one order of magnitude larger than the upper limit
       would be necessary to obtain a partial fitting with the observed SED
      (Yuan et al 2002). Furthermore, neither ADAF nor ADIOS models are able to
      reproduce reasonable SED-fitting with observations in the radio frequency regime.
      They seriously underestimate the radio luminosities in AGNs and black hole X-ray 
      transients\cite{Esin},\cite{DiMatteo}.
\item The dynamical range of ADIOS is largely overestimated. To clarify this point, let $r_\mm{tr}$ be 
      the transition radius, at which an SSD makes a transition to ADAF (Fig. 2).
      If both solutions are thermally and dynamically stable, why should an SSD change into
      ADAF at large radii, or vise versa?\\
      Mathematically: in the absence of external forces and for given Dirichlect boundary conditions,
      there is at most a single weak solution to the set of the corresponding linear systems of equations,
       which fulfils the entropy principle. Accordingly, at large radii the system of equations are nearly linear
      (i.e., they can be linearized as there are no mechanisms for generating shocks or strong gradients), 
      and the conditions imposed at the
      outer boundary determine the solution uniquely. Note that the equations describing
      accretion flows in SSD or ADAF are predominantly Eulerian, except the angular momentum and
      energy equations that contain second order partial differential  
      operators\footnote{In spherical geometry they read $L_\mm{E} =\nabla \cdot \lambda_\mm{FLD} \nabla E$, and
      $L_\mm{\ell}=\nabla \cdot \eta_\mm{tur} \nabla \Omega,$  respectively. E and $\lambda_\mm{FLD}$  are
      the radiation density and the radiative diffusion coefficient as defined in Eq. 12, 
      $\Omega$ and $ \eta_\mm{tur}$ the angular frequency and coefficient of turbulent diffusion,
      respectively.}.
 For example, if we set the angular velocity at the outer boundary to be
      sub-Keplerian and replacing $L_\mm{E}$ by $\Lambda_\mm{B}$ (see Eq. 12), the SSD changes to ADAF.
      However, since there is no reason to expect an excessive source of heating at large radii,
      as in ADAF case,   only cold standard disks are likely to survive there. These disks continue
      to maintain their thin geometrical structures all the way down to the central object, where
      other effects come into play, e.g., heat conduction,  change of the MF-topology,
     noticeable change of rotation-law and manifestation of the 
      magnetic field   and  the gravitational well of the central object.
      Indeed,  Abramowicz et al. (1998) concluded that a transition from SSD to ADAF may occur 
      in the vicinity of the BH. Hujeirat \& Camenzind (2000a) found that the transition is thermally
      unstable, and the plasma in the outermost part of ADAF collapses dynamically, thereby extending 
      the SSD down to several last-stable radii, where the disk truncates and an ion-dominated
      and highly edvective torus emerges.  \\
      An additional evidence that rules out transitions at large radii are MFs. ADAF/ADIOS models
      assume that MFs are in equipartition with the thermal energy. If this is the case, then this 
      should apply also when MFs are of large scale topology. However, since the accretion flow
      is fully ionized and freely falling, magnetic flux freezing implies that the MF-strength will
       strongly increase inwards and become in equipartition with the potential energy
      of the flow, hence terminating accretion. This impose a constrain on the location
      of the transition radius: $r_\mm{tr}$ must be smaller than $10-20$ last-stable radii, depending on
      the accretion rate.  
\een
\item Truncated disks - advective tori solution (TDAT, Hujeirat \& Camenzind 2000b).
      The model relies on the results of radiative HD calculations of two-temperature
      accretion flows onto BHs without magnetic fields. 
      These important results of TDAT read: 1) the disk truncate
      close to the last stable orbit, and forms an ion-dominated
      highly advective torus, 2) outward-oriented motions of plasmas between the
      disk and the overlying corona are formed, and which are manifested when MFs are
     included\cite{HujCL}.   In this paper, the previous model \cite{HujCL}
     is presented in details, and its astrophysical capability to reproduce the 
     spectral energy distribution of the jet in M87 is confirmed. 
\een


\section{Disk structure and magnetic braking  }
\subsection{The outer part}
In SSDs the inward-advection of angular
momentum
is balanced by outward viscous transport. This is equivalent to equalize the terms
T0 and T3 in Eq. 1, while all other terms are neglected. 

\beq
 \DD{\D \ell }{\D t}  + \overbrace{\nabla\cdot V \ell}^{\rm T0}
= \overbrace{B_\mathrm{r} \DD{\D r B_{\rm T}}{\D \mathrm{r}}}^{\rm T1}
  +  \overbrace{B_\theta \DD{\D B_T}{\D \theta}}^{\rm T2}
  +  \overbrace{
 \DD{1}{r^2} \DD{\D}{\D r}  r^4 \eta_\mm{tur} \DD{\D \Omega}{\D r}}^{\rm T3}
 + \overbrace{
 \DD{1}{\cos{\theta}} \DD{\D}{\D \theta}  \cos{\theta} \eta_\mm{tur}
 \DD{\D \Omega}{\D \theta}}^{\rm T4},
\eeq
where $\ell=r^2 \sin{\theta}\Omega$, $\eta_\mm{tur}=\rho \nu_\mm{tur}$ is 
the turbulent viscosity coeffecient \cite{Shakura}, $V$ is the velocity field
and $B$ is the magnetic field (see Eq. 8).
T0 denotes angular momentum transport via advection, T1 and T2 are for magnetic
extraction, and T3 and T4 are for viscous
(micro- or macroscopic hydrodynamical turbulence) re-distribution of angular
momentum. 

The key question here is how the constellation of these terms would look like
in the vicinity of the black hole.

In the outermost regions of a disk, we anticipate magnetic fields to be
below equipartition. 
Let $r_\mathrm{tr}$ be a transition radius (see Fig. 2), such that for 
$r > r_\mathrm{tr}$, we have
the usual SSDs , where the ratio of magnetic to gas pressure  is
$\beta  \ll1.$ Here the Balbus-Hawley instability 
\cite{Balbus}
operates on the dynamical time scale: it rapidly amplifies the MF and forces
$\beta$ to approach, but remains below, unity \cite{Hawley}. On the other hand, 
the rotational energy in SSDs exceeds the
thermal energy by at least one order of magnitude. Therefore, the generated toroidal  magnetic
energy  via shear  can easily exceed thermal equipartition. 
Consequently,  unless the imposed boundary conditions 
limit the amplification of the TMF, there is no reason to expect $\beta$ to remain strictly below  unity. 

\subsection{The inner part}
Whether the disk is surrounding compact object or a YSO, the innermost part of the
disk is likely to be strongly  magnetized. The question which then arise is:
how does accretion onto BHs proceed under $\beta \ge 1$ condition? 

In general, accretion via strongly magnetized and turbulence-free disks proceeds if the MFs are able to 
extract angular momentum from the rotating disk-plasma. The efficiency of this extraction
depends, among others, on the MF-topology. For example, large scale or dipolar
MF-topologies are considered to be appropriate for extracting rotational energy from the
disk  or even from the hole itself and power jets.
In the present study, MFs are assumed to be of large scale topology. This is reasonable, as the 
BH-instability in combination with the Parker instability may establish a dynamo cycle
in which the inward-advected and initially weak MFs are amplified and then reconnect to
build up the desired MF-configuration\cite{Tout}.
A straightforwards conclusion that can be drawn is that when the matter at the outer region has completed one revolution
around the central BH, MFs should have reached equipartition almost everywhere in the disk. 
This would eventually suppress self-generated turbulence and terminating 
angular momentum  transport through turbulence-friction.

In the region $r < r_\mm{tr}$, global conservation of the poloidal magnetic 
flux implies that
$B_\mm{P} \sim r^{-2}$.
Therefore, as the innermost part of the disk contracts, there is a critical radius $r_\mm{top},$
below which  gravitational-equipartition is maintained (i.e., Lorentz forces are of
 the same order as the gravitational forces), eventually terminating accretion.
This configuration is unlikely, as this would imply that most of 
accretion  inflow would be re-directed into outflow, with an unacceptably large mass load.
On the other hand, if $r_\mm{top}$ is sufficiently close to the last stable orbit, a change
of the MF-topology associated with a significant loss of magnetic flux through the horizon
would be possible. This is a plausible event, as MFs close to the horizon are unlikely to adopt
other than a monopole like-configuration (see Fig. 2/b).
Under such circumstances, the generation of $B_\mm{T}$ will be  considerably reduced and the 
accretion of matter within $r_\mm{top}$ proceeds  unhindered.
 
Around the radius $r_\mm{top}$ angular momentum transport is mediated through magnetic braking.
  When combining  
Eq.1 with the equation corresponding to the time-evolution of the 
TMF (see Eq. 8), and assuming the flow to be locally incompressible, we obtain a
magnetic torsional wave equation which has the approximate form: 

\beq
 \DD{\D^2}{\D^2 t}B_\mathrm{T}  \cong {V^2_\mathrm{A}}   
       \Delta {B_{\rm T}}.
\eeq
$\Delta$ denotes the two-dimensional Poisson operator in spherical geometry, and 
$V_\mathrm{A}$ ($\doteq B_\mathrm{p}/\sqrt{\rho}$)  is the {\Alfven} speed.
The action of these waves is to magnetic-brakes the innermost part of the disk through
transporting angular 
momentum from the disk to higher latitudes. Note that $B_\mathrm{p}$
 determines uniquely the speed of propagation, hence the  efficiency-dependence  
 of angular momentum transport on the $B_\mathrm{p}-$topology.
These torsional {\Alfven} waves (-TAWs) propagate in the vertical direction on the time scale:
\[ \tau_{\TAW} \sim \DD{H}{V_\mathrm{A}} \sim  \tau_\mathrm{dyn}. \]
 This should be similar to the removal time scale of angular momentum from the 
disk\footnote{$\tau_\mm{rem}$ is obtained from the terms T0 and T2 in Eq. 1}: 
 \beq
  \tau_\mathrm{rem} \sim \rho V_\mathrm{T} H/B_\mathrm{P} B_\mathrm{T} \sim r^{3/2}.
\eeq
Thus, $ \tau_\mathrm{rem}$ increases non-linearly with radius,  attaining a  minimum 
value at the inner boundary.
This implies, for example, that the rate at which angular momentum is removed
at the last stable orbit ($\doteq R_\mathrm{LSO}$) is one thousand times faster
than at  $r=100 R_\mathrm{LSO}$. To avoid the innermost region
of the disk from running out of angular momentum, we require the disk to be dynamically stable.
Equivalently,  the rate at which angular momentum is removed at any radius  must be 
equal to the rate at which it is advected
inwards from outer layers, i.e., $\tau_\mathrm{adv}= \tau_\mathrm{rem}$. 
This implies that the radial velocity
$-U$  in the disk is of the order of the {Alfven} speed, which is  also of the order
or even larger than the sound speed $V_\mathrm{S}$, hence the terminology advection-dominated disk. \\

The question which arises here is whether the disk, under these circumstances, maintains
its geometrical thin structure.

The answer is hidden in the vertical momentum and energy equations. The former reads:  
\beq
  \DD{\D V }{\D t}  +  V \cdot \nabla V =  \DD{1}{\rho r} \DD{\D P}{\D \theta}
  + \tan{\theta} \DD{V^2_\mm{\varphi}}{r} + \DD{\nabla \times \nabla \times B}{4 \pi \rho}|_{\vec{\theta}}.
\eeq
Taking into account that vertical transport of angular momentum is 
maintained through magnetic braking nearly without particle-advection
(i.e.,  we neglect the second term in the LHS of the equation), and noting that 
$({\D P}/{\D \theta})/{r \rho}$ is the only term that opposes the vertical 
 contraction of the disk,  we obtain:
\beq
 \DD{H}{r} \approx \DD{V_\mm{S}}{V_{\varphi}}.
\eeq
This implies that, as in SSDs, the thickness of the disk depends strongly on the temperature of the plasma, and hence 
on the associated heating mechanisms. On the other hand, since $\beta \ge 1$ in the innermost part of the disk,
the dominant source of heating in SSDs( i.e., turbulence dissipation), is no longer efficient. In this case,
heating is mainly due to adiabatic compression and to other non-local sources, e.g., radiative reflection, 
Comptonization and conduction of heat flux from the surrounding media. However, these mechanisms
are unlikely to change the disk configuration significantly, so that the disk continues to  
maintain its thin geometrical structure.  Specifically, for $ r \le r_\mm{tr}$, we
have $\D H_\mm{d}/\D r \ge 0$.

\subsection{Formation of the ion-dominated torus}
Whether the disk model is an SSD, ADAF or ADIOS,  most of the physical variable (e.g.,  density
temperature, radial and angular velocities) increase with decreasing radius.
On the other hand, the disk is said to be truncated if within a certain radius,
say the radius of the boundary layer $r_\mm{BL}$, part of the main physical  variable start to decrease inwards. 
This occurs, for example, if the central object rotates sub-Keplerian\footnote{This applies
for Kerr black holes with $a<1$.}. The Kepler-rotating and inflowing material from
the disk must release a significant fraction of its angular momentum to be able to cross the event horizon of a
Schwarzschild BH. This might be achieved if the MFs in the BL are sufficiently strong
to extract angular momentum from the disk on the dynamical time scale. However, this requires
the density to decrease considerably inwards, so that the speed of propagation
of the TAWs largely exceeds that in the outer disk.

We note that, unlike SSD and ADAF-models that predict a density-decrease with radius 
(i.e, $\rho \sim r^{-15/8}$ and $\rho \sim r^{-3/2}$, respectively), hydro- and
MHD-calculations show, indeed, an inwards decrease of the  density in the BL\cite{HujCa},\cite{Stone}, \cite{Popham}.\\
In SSDs the pressure gradient $\nabla P_\mm{gas}$ is neglected, whereas in ADAF 
the pressure adopts the profile $p\sim \sim r^{-5/2}$. When comparing 
the gravitational to the advective time scale at the inner boundary, 
it is easy to find that the rate of advection increases
inwards, forcing thereby the density to be minimum at the inner boundary. As a 
consequence, the in-flowing material at $r_\mm{BL}$ experiences a pressure-induced acceleration.
 Taking into account that the Coulomb coupling
between the ions and electrons is $\Lambda \propto \rho^2$, thermal decoupling may occur, and
 formation of an ion-dominated torus is an inevitable phase in the evolution of accretion flows
  onto BHs, at least for sub-Eddington accretion. 

How does heat conduction affect the structure of the ion-torus?\\
Generally, the temperature of both the ions and electrons increase inwards. 
The effect of heat conduction is to transport heat from hot into cold plasma, i.e, from inside-to-outside. 
In the absence of magnetic field and assuming the mean turbulent motion to be proportional to
the sound speed, the innermost part of the disk will be evaporated by heat conduction
of the ions,  and therefore the torus starts to expand outwards \cite{HujCb}.
 When large scale MFs 
are taken into account, the torus contracts rather than expands.  The reasons are:
1) heat conduction operates parallel to the MF-lines, and 2) strong MFs act to diminish 
  the generation and dissipation of turbulence.\\
Unlike stars that heat up the surrounding media,  BHs are heat-absorber (the surrounding
corona is not heated from below). Therefore, the ion-dominated torus contracts
and may survive in the vicinity of the last stable orbit only, where MFs are 
 pre-dominantly of large scale topology.

In addition, the location of  $r_\mm{tr}$ depends strongly on the accretion rate $\Mdot$.
A large  $\Mdot$ enhances the Coulomb coupling, reduces the effect of heat conduction
and thereby slowing the  propagation speed of the TAWs (which reduces the efficiency of
 the magnetic braking). Thus, the volume of the ion torus shrinks and the total power
injected into the jet-plasma decreases accordingly.\\
\begin{figure*}
\begin{center}
{\hspace*{-0.5cm}
\includegraphics*[width=12.0cm]{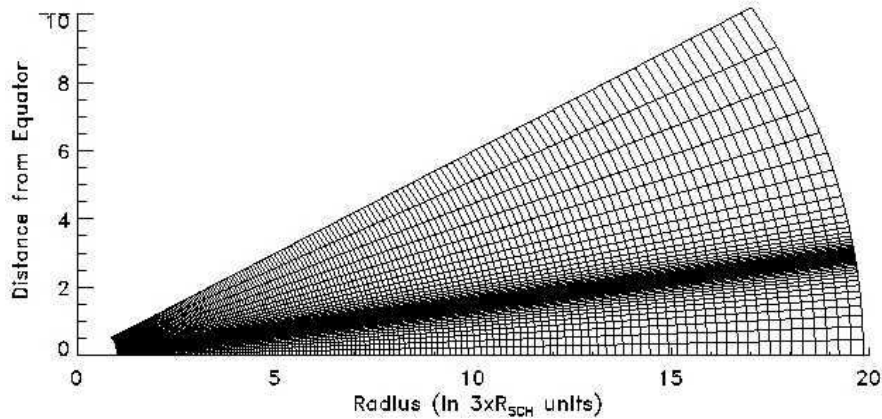}
}
\end{center}
{\vspace*{-0.4cm}}
\caption [ ] {A partial display of the distribution of 220 and 100 finite
  volume spherical cells in the radial and 
              horizontal directions that covers the domain of calculations.
        Highly refined mesh distribution is adopted 
          in the transition layer, where most of the
              physical variable  undergo a strong spatial variation and where 
               jet-launching occurs. The aspect ratio adopted here, i.e., 
              the ratio of the finest volume cell to the largest one, 
                  is $\approx 10^4 $.  
  } 
\end{figure*}
\section{The governing equations}

When modelling flow-configurations in the vicinity of a BH,
it is necessary to use spherical geometry. Since the
dynamical time scale close to the last stable orbit is extremely short,
thermal decoupling between the electrons and the protons is possible.
In this case,  the set of equations to be solved consists of the continuity equation:
\beq
   \DD {\D \rho}{\D t} + \nabla \cdot (\rho \vec{V})  =  0,
\eeq

the three momentum equation for the material flux: 
\beq
  \DD {\D \vec{M}}{\D t} + \nabla \cdot(M \otimes \vec{V_i}) = -\nabla P_{\rm g} 
  + {\bf {f}}^{grav,cent,rad} + \DD{\nabla \times B \times B}{4 \pi} + Q^{\rm visc}, 
\eeq
where $\rho,~\vec{V}=(V_r,V_\theta,V_\varphi),~\vec{M}$ and P are the density, velocity vector,
 material flux $\vec{M}= \rho \vec{V},$ and
gas pressure $P (\doteq  {\cal R}_{\rm gas} \rho (T_i/\mu_{\rm i} + T_e/\mu_{\rm e})).$  The subscripts
``i'' and ``e'' correspond to ions and electrons. $\mu_{\rm i}=1.23$ and $ \mu_{\rm e}=1.14 $.
 ${\bf {f}}^{grav,cent,rad}$
 is the force vector which includes the gravitational,
centrifugal and radiative forces. 
Quasi--Newtonian gravity is used to describe the gravity of the central BH \cite{Paczyanski}.
$Q^{\rm visc}$ denotes the collection of second order diffusive
operators.\\ 
The evolution of the MFs is followed by solving the induction equation
 which reads: 

\beq
        \D_t\vec{B} = \nabla\!\times\!
        (\vec{V}\!\times\!\vec{B} - \eta_\mm{mag} \nabla\!\times\!\vec{B}) .
\eeq
Here, $\vec{B}=(B_{r},B_{\theta},B_{\varphi})=(B_\mathrm{p},B_\mathrm{T}).$
To quench the amplification of the TMF in the TL, we adopt  
the magnetic turbulent diffusivity: $\eta_\mm{mag} = \alpha V_\mathrm{AT} H_\mathrm{d}\,,$
where $ V_\mathrm{AT} = B_\mathrm{T}/\sqrt{\rho}$.  
In solving this equation for $\,\vec{B}\,$, the solenoidal condition
$\nabla\cd\vec{B}=0\,$ must be satisfied everywhere and for all times
\cite{HujR}.

The dynamical time scale in the innermost region of the disk is sufficiently
short, that
thermal decoupling between the electrons  and ions is inevitable \cite{Rees},  \cite{Narayan}. To take this possibility into account, both equations describing the
thermal evolution of the  ions and  electrons should be solved. In this formulation,
the bulk of heat generated by turbulent dissipation goes into heating mainly 
the ions in virtue of their large  mass compared to the electrons. 
The dominant cooling mechanisms for the ions are conduction and two-body
interaction with the electrons.
The electrons, on the other hand, 
 are subject to various cooling processes, e.g.,  Bremsstrahlung,
Comptonization, Synchrotron and conduction. Surprisingly, ion-conduction appears 
to play an important role in the innermost region of the disk, where the ions
form an ion-dominated torus and evaporate the inner part of the disk 
\cite{HujCb}.  
\begin{figure}[htb]
\begin{center}
{\hspace*{-0.5cm}
\includegraphics*[width=7.5cm]{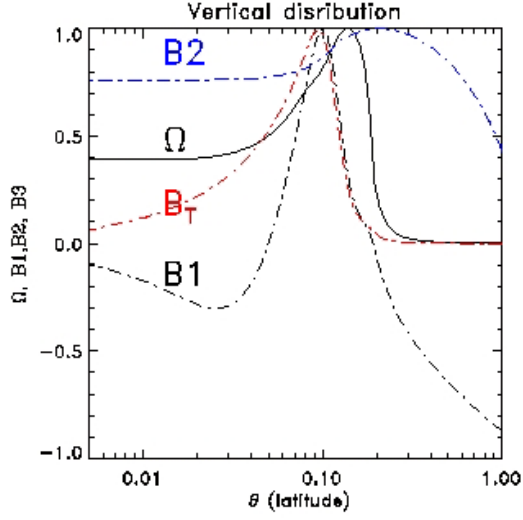}
}
\end{center}
{\vspace*{-0.4cm}}
\caption [ ] { The horizontal distribution of the normalized   
angular velocity $\Omega$,  the poloidal magnetic field components $(B1, B2)=(B_\mathrm{r}, B_\theta)=B_\mathrm{P}$
 and the TMF ($B_\mathrm{T}$) at $r=2.5$. Note the  super-Keplerian rotation and
the strongly enhanced strength of the MF-components in the TL.   } 
\end{figure}
 

\begin{figure*}[htb]
\begin{center}
{\hspace*{-0.5cm}
\includegraphics*[height=8.0cm,width=15.5cm]{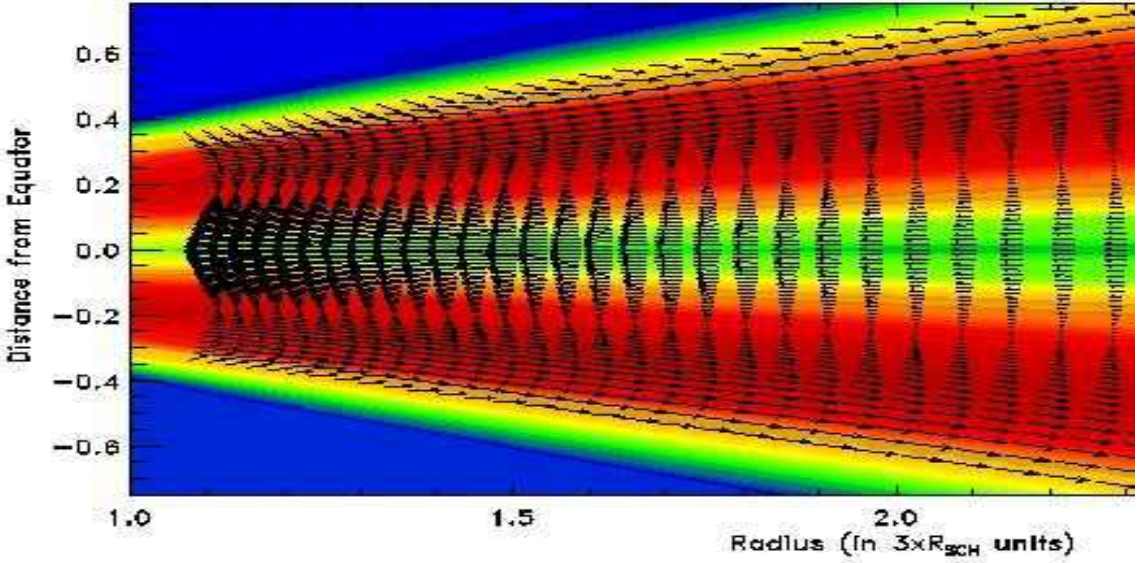}
}
\end{center}
{\vspace*{-0.4cm}}
\caption [ ] { The distribution of the velocity field in the innermost part of the disk,
               superposed on the logarithmic-scaled  ratio of the ion- to electron-temperatures
            (red color corresponds to high ratios
           and blue to low-ratios).   } 
\end{figure*}
 
Under these conditions, the respective internal energy equations of the ions and electrons read:
 \beq
 \DD { \D {\cal E}^i}{\D t} + \nabla \cdot ( {\cal E}^i  \vec{V}) = 
         P^i (\nabla \cdot \vec{V}) + \nabla \cdot [\kappa^i_{\rm cond} \nabla T_{\rm i}]
            + {\cal{D}} - \Lambda_{i-e},
\label{EQcondi}
\eeq
 \beq
 \DD { \D {\cal E}^e}{\D t} + \nabla \cdot ( {\cal E}^e  \vec{V}) = 
         P^e (\nabla \cdot \vec{V}) +\nabla [\cdot \kappa^e_{\rm cond} \nabla T_{\rm e}]
         +  \Lambda_{i-e} - \Lambda_B  - \Lambda_{C} - \Lambda_{syn}
\label{EQconde}
\eeq

where $ P^{e,i} = {\cal R}_{\rm gas} \rho (T_i/\mu_{\rm i} , T_e/\mu_{\rm e}),$  
${\cal E}^{e,i}= P^{e,i}/(\gamma-1)$ and $\gamma=5/3.$
 ${\cal{D}}$, $\Lambda_{B}$, $\Lambda_{\rm i-e}$, $\Lambda_{\rm C}$, $\Lambda_{\rm syn}$ are
the turbulent dissipation rate, Bremsstrahlung cooling, Coulomb coupling between the ions
and electrons, Compton and synchrotron coolings, respectively. Here, we use the following express:
 \[\Lambda_{\rm i-e} = 5.94\times 10^{-3} n_i n_e c 
      k \DD{(T_{\rm i} - T_{\rm e})}{T^{3/2}_{\rm e}} \]
\[ \Lambda_{\rm B} =  4ac \kappa_\mathrm{abs} \rho (T^4 - E), \]
\[ \Lambda_{\rm C} = 4 \sigma n_e c  (\DD{k}{m_e c^2}) (T_e - T_{rad})E, \]
 
where $\kappa_\mathrm{abs}$ and $\sigma$ are the absorption and scattering  coefficients.
$n_{\rm e},~n_{\rm i}$ the electron- and ion-number densities. The 
radiative temperature   is defined as $T_{rad}=E^{1/4}$, and E is
the density of the radiative energy in the zero-moment approximation of
the radiative field.
For the  synchrotron cooling function, the following relation is adopted\cite{Rybicki}:
\beq
  \Lambda_{\rm syn} = \DD{E_\mathrm{mag}}{E} \Lambda_\mathrm{C},
\eeq
where $E_\mathrm{mag},\,E$ are the magnetic and radiative 
energies, respectively.

The second order operators on the RHS of Equations (\ref{EQcondi}) and  (\ref{EQconde})
  are conductive operators and the
conduction coefficients in cgs units read (\rm{Sandb{\ae}k}\,\&\,Leer 1994): 
 \[\kappa_\mathrm{e} = 7.8\times 10^{-7}\,T_\mathrm{e}^{5/2}, ~~
 \kappa_\mathrm{i} = 3.2\times 10^{-8}\,{\rm T}_\mathrm{i}^{5/2}\]

To follow the evolution of the radiative field, several assumptions have been made.
Mainly, it is assumed that the field is isotropic and that the Flux-limited-Diffusion (FLD) approximation
can be used to close the set of radiative moments. Thus, only the 0-moment equation of
the radiation field is solved.
Since the optical depth in the disk can be large in certain regions, and small in others,
FLD approximation is used to model the radiative flux appropriately  in these different regions, and
to assure the monotonicity of the flux in the transition regions in-between. The 0-moment of the radiation field
then reads: 
\beq
 \DD { \D {E}}{\D t} + \nabla \cdot ( {E}  \vec{V})  = 
         \nabla \cdot [\lambda_\mathrm{FLD} \nabla {E}]
               - \Lambda_B + \Lambda_{C} + \Lambda_{syn},
\eeq
where $\lambda_\mathrm{FLD}$ is 
flux limited diffusion coefficient which forces the radiative flux to adopt
the correct form in optically thin and thick regions, i.e.,  
\beq
{\nabla  \cdot \DD{\lambda_{\rm FLD}}{\chi}\nabla E = 
        \left\{ \begin{array} {c@{\quad \quad}l}
  \nabla \cdot \DD{1}{3\chi}\nabla E  & {\sf if} \hspace*{0.75cm} \tau \gg 1  \\
  \nabla \cdot nE  & {\sf if} \hspace*{0.75cm} \tau \ll 1,  
        \end{array} \right. }
\eeq
and provides a smooth matching in the transition regions. Here
 $\chi = \rho(\kappa_\mathrm{abs} + \sigma)$ and ${\rm n = \nabla E/|\nabla E|}.$
\subsection{Method of solution, initial and boundary conditions}

The solution procedure is based on using the implicit finite volume solver IRMHD3
 to search steady-state solution for the above-mentioned 3D axi-symmetric, 
two-temperature, diffusive,
radiative and  MHD equations in conservative form \cite{HujR}.

We note that the strong non-linearities governing these equation may not admit
steady state, but strongly time-dependent or, in the best cases, quasi-stationary
solutions. This implies that the terms describing the time-variation of the
variables, i.e., $ \D/\D t,$  must be retained. On the other hand, since
the dynamical time scale for accretion flows is shortest at the inner boundary,
searching for quasi-stationary for the global flow-configuration on these time
scales could be a computationally prohibitive process.
 
To overcome this problem, we use the following spatially varying time stepping (-SVTS)
approach, in which 
\beq
   \delta t_\mm{i} = \alpha\,\delta r_\mm{i},
\eeq
where $\delta t_\mm{i}, ~\delta r_\mm{i} $  are the time step size and the radial mesh increment
of a finite volume cell at a given radius $r_\mm{i}$. The coefficient $\alpha$ is a constant of
order unity.\\
This SVTS method is appropriate for searching quasi-steady configuration of accretion flows
or eveb for the collapse of cloud cores in molecular clouds, in which the radial grid points are
non-linearly distributed and $\delta r$ increases with radius. 
If  $\delta r$ is constant, then $\alpha$ must depend on the radius. In this case
we suggest $\alpha = \alpha_0 (r/r_\mm{in})$, where  $\alpha$ is a constant of
order unity and $r_\mm{in}$ is the radius at the inner boundary.\\
The main disadvantageous of SVTS method is its inability to provide time scales
for features that possess quasi-stationary behaviour, and that might be of astrophysical
relevance. Here we suggest to use the obtained quasi-stationary
solutions as initial configuration and re-start the calculations using
a uniform time step size.

We note that SVTS method can be applied to explicit numerical methods as well.
Here  the global Courant number should be replaced by a local one, so that
each finite volume cell has its own time step size. \\ \\
In the present paper, the equations are solved in non-dimensional form, using the 
reference scaling variables:
$\rm{\tilde{\rho}= 2.5 \times 10^{-12} {\rm g\, cm^{-3}}, \tilde{T}= 5 \times 10^7 K},$
$\tilde{U}=\tilde{V_\mathrm{S}} = \gamma {\cal R}_\mathrm{g} \tilde{T}/
\mu_\mathrm{i},$ $(\mu_\mathrm{i}=1.23)$.
$\tilde{B}=\tilde{V_\mathrm{S}} \sqrt{4 \pi \tilde{\rho}}.$ 
The location of the transition layer (-TL), where the ion-dominated plasma
is expected to rotate super-Keplerian and centrifugal-accelerated into jets,
is shown in Fig. 2 for clarity.

The central object is taken to be a $10^8M_{\odot}$ Schwarzschild BH.
The domain of calculations is the quadrant:  
$D=[0\le \theta \le \pi/2] \times [ 1\le {r}\le 20]$. Radii 
are given in units of the inner radius
 $R_{\mathrm{in}} \doteq 2.8\,R_{\mathrm{S}}$, where 
 $R_{\mathrm{S}}$ denotes the
Schwarzschild radius $R_{\mathrm{S}} \doteq 2GM/c^2$.  
Through the outer boundary and within the thickness
 $H_\mathrm{d} = 0.1 r,$ the constant accretion rate  $\Mdot = 0.01\times \Mdot_\mathrm{Edd}$  
is assumed. The inflow-temperature is taken to be $T=10^{-3} T_\mathrm{virial}$. The ion-temperature
  $\rm T_\mathrm{i}$  is set to be equal to the electron temperature 
$\rm T_\mathrm{e}$ initially. 
A hot and tenuous corona
($T=T_\mathrm{virial}$, and density $\rho(t=0,r,\theta)=10^{-4}\rho(t=0,r,\theta=0$)
is set to sandwich  the disk.
Initially, the MF is set to be of large scale topology and in thermal equipartition with the
disk-plasma at the outer boundary. 
Across the inner boundary, free-fall for the radial velocity and stress--free conditions for
the angular velocity are imposed. Normal symmetry and anti--symmetry
conditions are assumed along the equator and along the polar axis. The
domain of integration is divided into $220\times 80$ strongly stretched finite
volume cells in the radial and vertical directions, respectively (Fig. 3).
  
Special attention was given to the physical consistency of the imposed 
conditions that MFs must fulfil at the inner and outer boundaries.
This is especially important in magnetic-diffusive plasma, as
second order magnetic diffusive operators may allow informations to be communicated
between the inner and outer boundaries on the dynamical time scale. Therefore, 
magnetic stress-free conditions have been imposed at both boundaries.
This is equivalent to require $\nabla \times B = 0$. Thus, the 
magnetic diffusive operators vanish at the inner and outer boundaries, which is a reasonable
assumption for accretion-disk inflows around BHs, in which the magnetic Reynolds numbers
can be extremely large in certain regions and small in others.
\begin{figure*}[htb]
\begin{center}
{\hspace*{-0.5cm}
\includegraphics*[width=13.5cm]{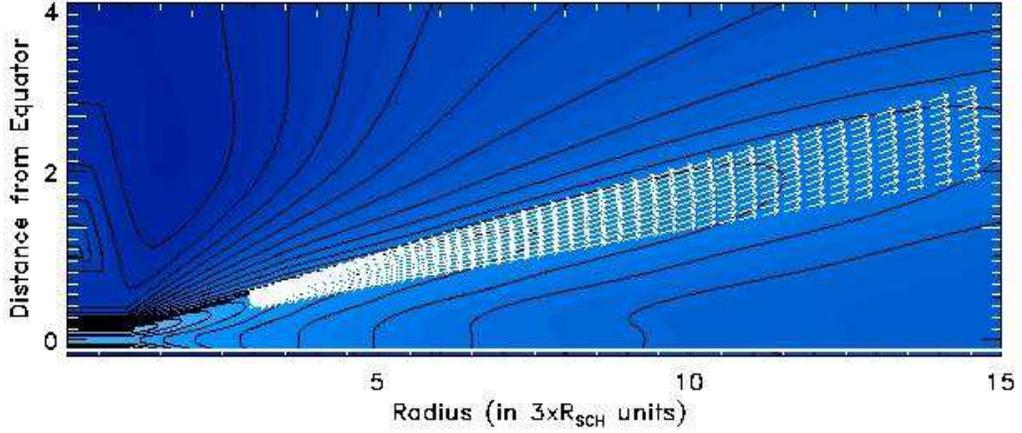}
}
\end{center}
{\vspace*{-0.4cm}}
\caption [ ] { The distribution of the velocity field in the transition region 
           superposed on equally-spaced isolines of the
          angular velocity $\Omega$. The strong-decrease of $\Omega$
          with radius in the equatorial region relative to its slow-decrease in the
          TL is obvious.  
  } 
\end{figure*}

\begin{figure*}[htb]
\begin{center}
{\hspace*{-0.5cm}
\includegraphics*[width=13.5cm]{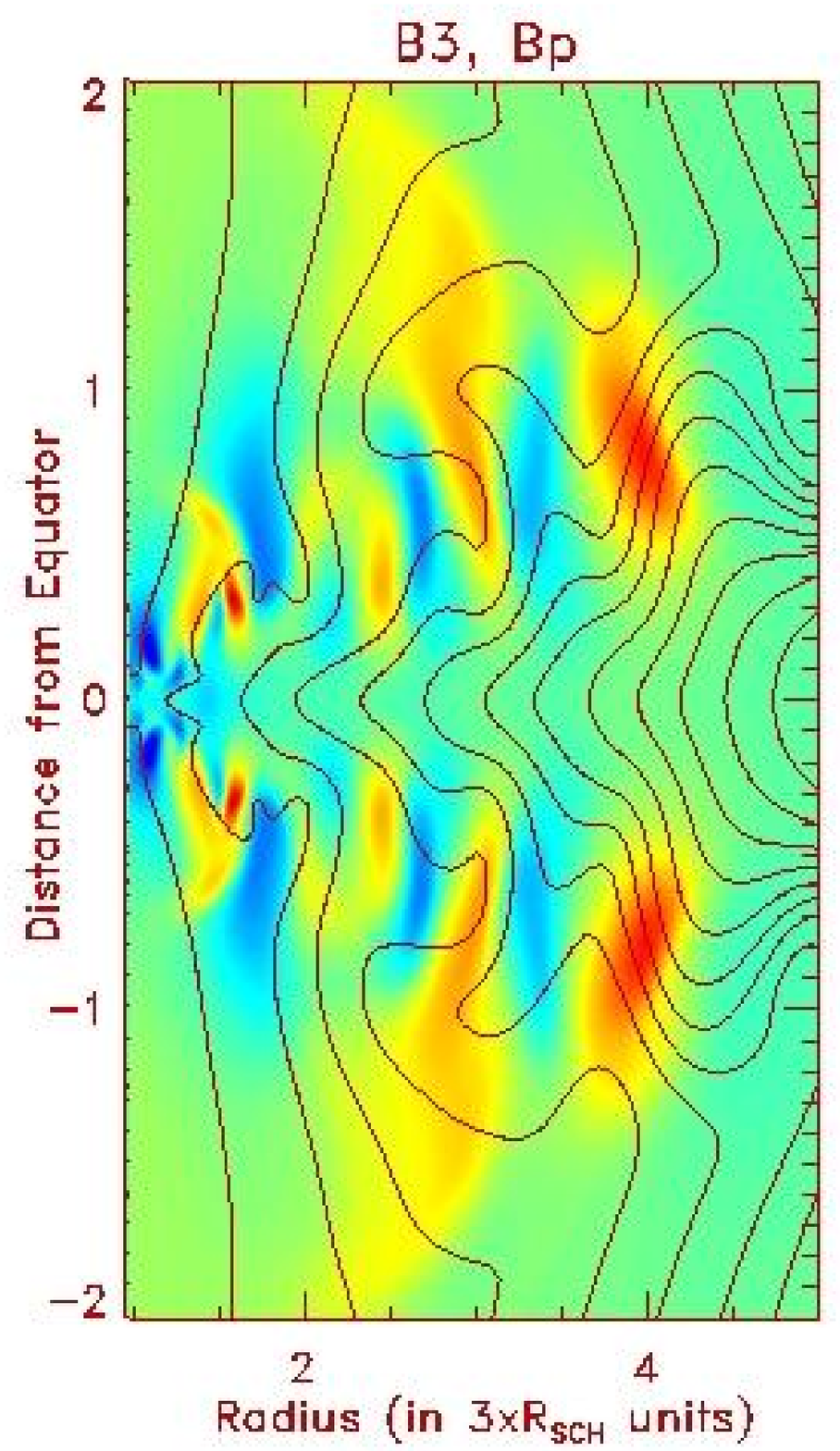}
}
\end{center}
{\vspace*{-0.4cm}}
\caption [ ] { The distribution of 30 weak poloidal magnetic field lines around a non-rotating
                BH after 10 orbital periods at the inner boundary (black lines).
               The TMF-distribution  is shown as well (blue color corresponds to large positive 
               values, and red to negative  values). In this run we set $\beta=1/20$, the disk is
               relatively hot ($T(\theta=0,r=r_\mm{out} = 10^{-1} T_\mm{virial}$)
               and thin ($\Mdot = 10^{-3}\Mdot_\mm{Edd}$). 
               Obviously,  the poloidal and toroidal MF components are strongly
               time-dependent and attain maximum values in the TL (to be compared with 
                Fig. 4 in \cite{HujCL}, where an initially large $\beta-$inflow 
               is assumed).} 
\end{figure*}

\section{Results: formation of the super-Keplerian layer}
When an isolated Keplerian-rotating particle in the equatorial near-plane is shifted to
higher latitudes while conserving its angular momentum, its rotation becomes super-Keplerian
locally, and therefore it starts to move outwards. Moreover, noting that $\tau_\mathrm{TAW}$ 
decreases strongly inwards, a transition layer between the disk and the hot tenuous corona
starts to form from inside-to-outside, where the matter rotates super-Keplerian.
Based on a previous calculations \cite{HujCL}, 
it was found that:   1)  the angular velocity  in the TL
 adopts the approximate power law profile
$\Omega \sim r^{-5/4}$, 2) ions cool predominantly through fast outflows,  and 
3) the amplification of the generated TMF is equilibrated through outward advection and
   magnetic-quenching
   (i.e., we artificially increased the coefficient of the magnetic diffusion in the TL
   to allow a flux-loss of the TMF). 
 In the TL, one possible solution for the radiative MHD equations is a self-similar
 solution in which the variables obey the following power-laws:
\beq
   \Omega   \sim r^{-5/4},
   \rho   \sim r^{-3/2},
   T_\mathrm{i}  \sim r^{-1/2},
   U_\mathrm{r}  \sim r^{-1/4},
   B_\mathrm{P}/B_\mathrm{T} \sim {\rm const.} = \epsilon, 
\eeq
where $\epsilon=H_\mm{W}/r,$ and $H_\mm{W}$ is the width of the TL.
The most prominent properties of this solution are:
\ben
\item The ratio of heating to advection time scale of the ions is of order unity, i.e., 
         $ \tau_\mm{h}/\tau_\mm{adv} \approx 1$. The ratio of heating to Coloumb-cooling
         time scale of the ions is:  $ \tau_\mm{h}/\tau_\mm{Coul} \sim r^{-1/2}.$ 
         This implies that thermal coupling between electrons and ions diverges
         with distance from their origin, and gives rise to multi-component jet-plasma.
         Furthermore, ions cool pre-dominantly through
         advection, i.e., adiabatic expansion.
\item   The ratio of the amplification of the TMF to the flux-loss time scale is of order unity.
        The same applies to the ratio of the advection to amplification time scale.
       Thus, this implies that a significant fraction of the toroidal magnetic flux is
        advected outwards, and which thereafter acts to collimate the flow. On the other hand, 
        the magnetic diffusivity in the TL scales: 
      $\eta_\mm{mag} \approx H_\mm{W}(\Bp/\Bt)V_{\varphi} \sim \epsilon r^{3/4}.$
     This $\eta_\mm{mag}-$ description must break down at a certain radius $r_\mm{col}$, where
      $\eta_\mm{mag}$ becomes negligibly small and  collimation starts to be efficient.  
\item 
       The rate of the outward-oriented material flux in the TL 
reads:
\beq
   \Mdot_\mm{W}(r) = \Mdot_\mm{W}(r=r_\mm{in}) + {\epsilon^2} \Mdot_\mm{d}(r=r_\mm{in})
 ( (\DD{r}{r_\mm{in}})^{1/4} -1),
\eeq
 where $\Mdot_\mm{W}(r=r_\mm{in})$ and $\Mdot_\mm{d}(r=r_\mm{in})$ are the wind-flux rate
 and the disk accretion rate at the inner boundary.
In writing Eq. 16 we have r-integrated the continuity equation:
 $\D{\Mdot_\mm{W}}/\D r = H_\mm{d} (\rho V)|_{\theta_\mm{d}} = \epsilon^2 r^{-3/4}$,
using $H_\mm{d} = r \sin{\theta_\mm{d}}$, $\epsilon = H_\mm{d}/r$ and $V = (H_\mm{d}/r) U.$
$\theta_\mm{d}$ corresponds to the latitudinal angle between the equator and the surface of the disk
is spherical geometry.

 Thus, the wind flux increases with 
 radius, although it is a rather weak dependence. 
\item the associated angular momentum flux  with the wind increases linearly
      with distance from the central BH: 
\beq
\dot{\cal J} = \rm{const.}\,r,
\eeq
which applies for $r\le r_\mm{tr} \approx 10-20\, R_\mm{LSO}.$
\een
The global energy distribution in the disk and jet region can be well represented 
by the  Bernoulli number (Be). Flows with total positive energy are gravitationally unbound,
and potentially they may propagate to infinity, whereas flows with negative total energy
are gravitationally bound and they end their motion inside the BH. If the flow, however,
is dissipative, energy exchange between different parts of the flow is possible,which
gives rise to inflow-outflow configurations.  
Fig. 9, for example shows that Be is  everywhere negative safe
the TL, where it attains large positive values. 
\begin{figure*}[htb]
\begin{center}
{\hspace*{-0.5cm}
\includegraphics*[width=13.5cm]{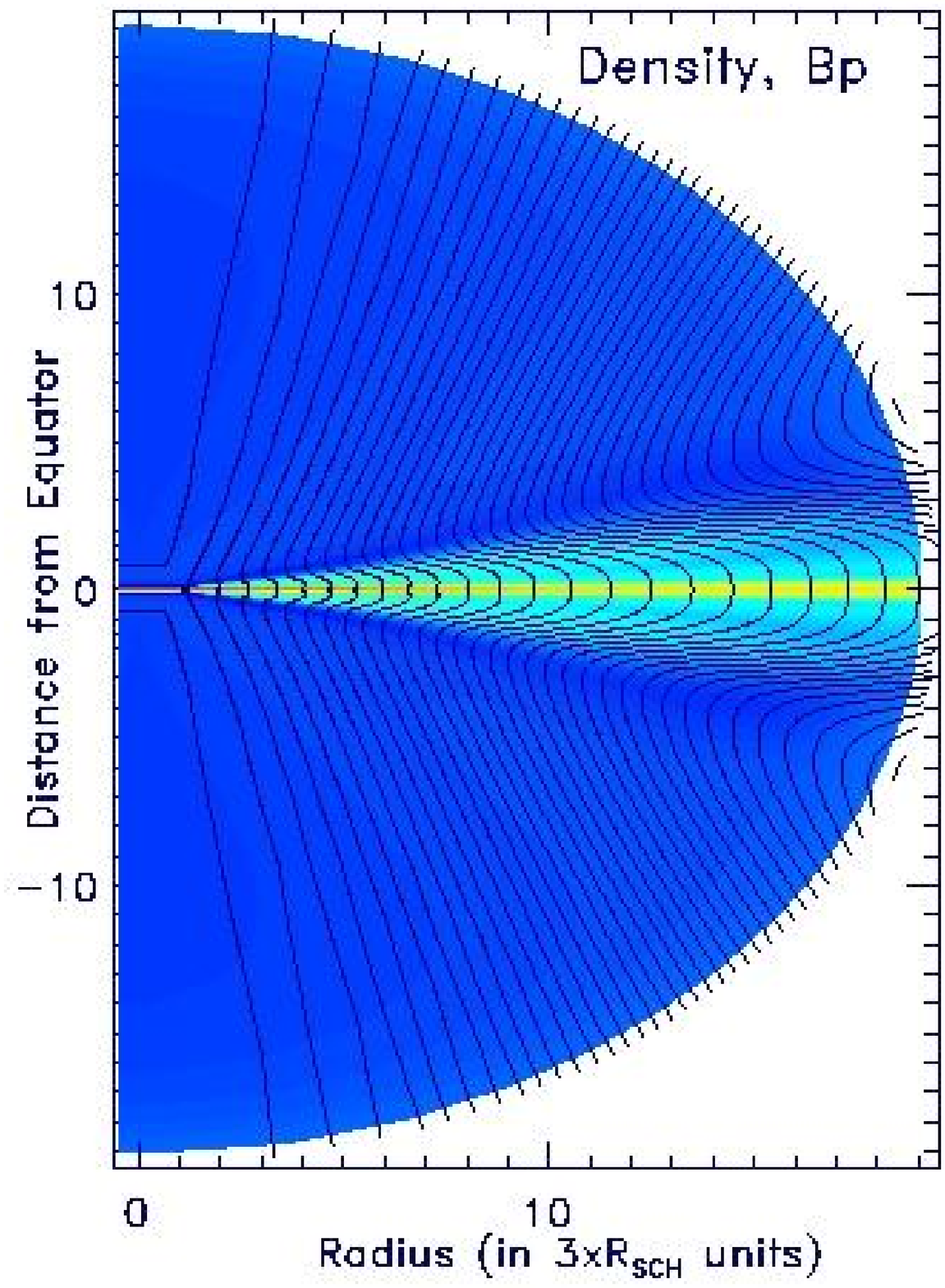}
}
\end{center}
{\vspace*{-0.4cm}}
\caption [ ] { 30 equally-spaced isolines of the poloidal-component $ \rm B_\mathrm{P}$
 ( black lines) and
               the density-distribution (yellow color corresponds to
               very high density-values, blue to middle and violet to low values).  } 
\end{figure*}
 
 
\begin{figure*}[htb]
\begin{center}
{\hspace*{-0.5cm}
\includegraphics*[width=13.5cm]{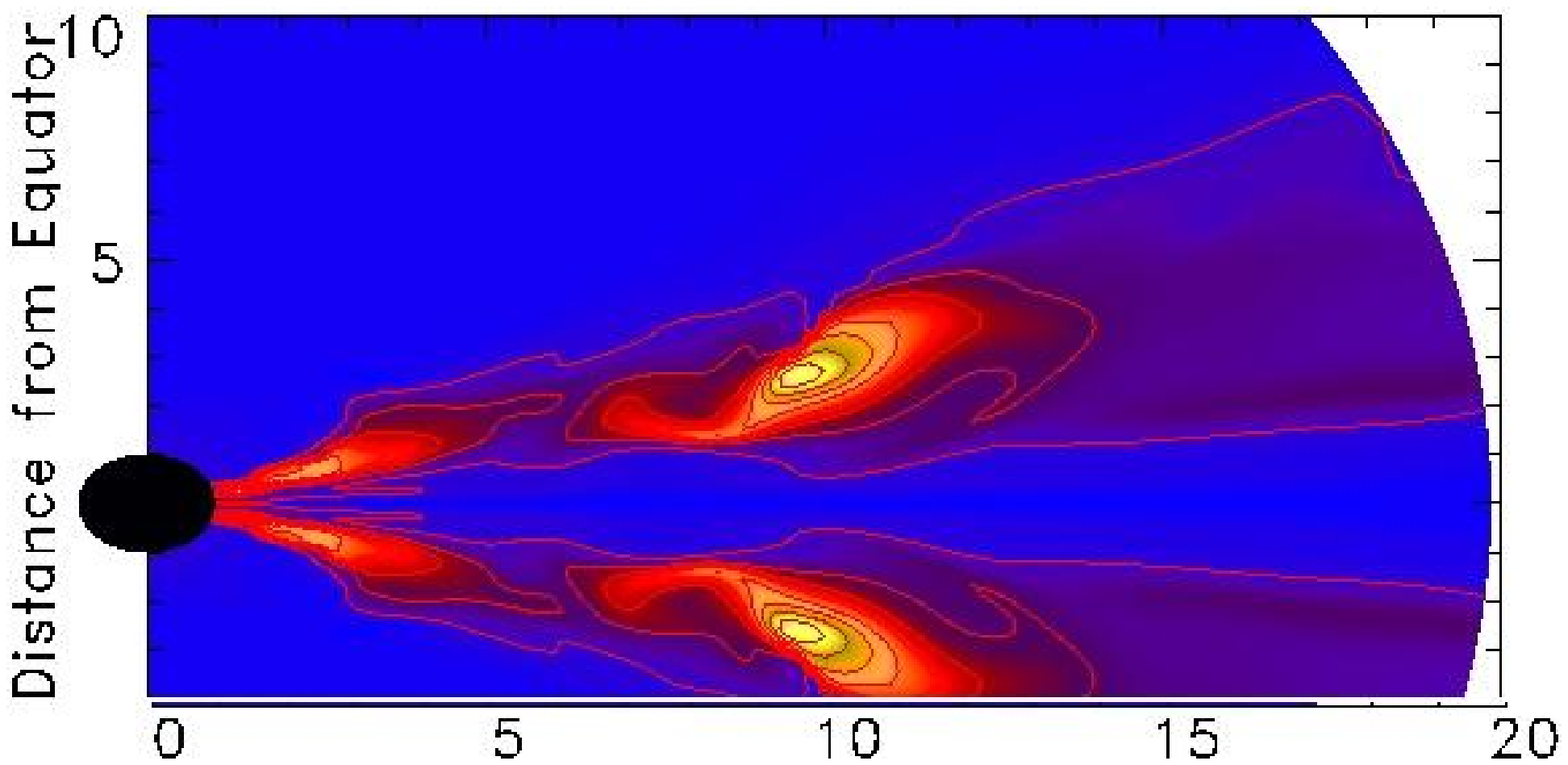}
}
\end{center}
{\vspace*{-0.4cm}}
\caption [ ] { A snap-shot of the distribution of the Bernoulli number in 
               two-dimensional quasi-stationary calculations. 
              The decrease from large to low positive values is represented
              via yellow, green and red colors.  The blue color corresponds
              to negative values.  The figure shows two ejected gravitationally unbound 
              blobs of large positive energies in the TL.    } 
\end{figure*}
 

We note that the outflow is sufficiently strong to shift the poloidal magnetic field (-PMF)
 lines outwards, 
whereas the large magnetic diffusivity prevents the formation of large electric currents
along the equator. In the case of weak MFs ($\beta \le 0.1$),  
our calculations reveal rather weak outflows, which is reasonable, since a low $\beta $ yields
a longer $\tau_\mm{TAW}$, which may become longer than the dynamical time scale.
Moreover, low $\beta$s inflows force MFs to establish a monopole
like-configuration (i.e, a one-dimensional MF-topology which is not appropriate for
producing strong TMFs). In this case, the rate of TMF-generation is considerably reduced. 
TAWs become inefficient at magnetic-braking the disk, reducing therefore the power 
required for launching jets. This indicates that weakly magnetized disks
are in-appropriate for launching  powerful jets, and that magnetized accretion flows
are more suited for jet-production than their HD-counterparts. 
Comparing the flux of matter in the wind region to that in the disk, it has been found that
${\Mdot}_\mathrm{W}/{\Mdot}_\mathrm{d} = {\rm const.}\sim \epsilon^2$. The angular momentum flux associated with
the wind  is
 $\dot{\cal J}_\mathrm{W}/\dot{\cal J}_\mathrm{d} = {\rm a }\, 
({\Mdot}_\mathrm{W}/{\Mdot}_\mathrm{d})\,r^{1/4},$ where ${\rm "a" }$ is a constant of order unity.
Consequently, at 500 gravitational radii almost $15\%$ of the total accreted angular momentum
in the disk re-appears in the wind. 
To clarify why the TL is geometrically thin, we note that the ratio:
\beq
\DD{\tau_\mm{TAW}}{\tau_\mm{adv}} \sim (\DD{r}{H_\mm{d}}) (\DD{V_\mm{A}}{U}) \sim  \DD{r}{H_\mm{d}}
 > 1,
\eeq
where $\tau_\mm{adv}$ is the advection time scale. The last inequality implies that 
angular momentum in the TL will be advected outwards more efficiently than being
extracted through TAWs to higher latitudes. Furthermore, since the flow in the TL is highly
dissipative, TMF-lines tend to close in the TL, thereby considerably reducing the efficiency
of angular momentum transport to much higher latitudes.
Also, as the flow in the TL rotate super-Keplerian, the normal component of the centrifugal
force tend to shift the matter towards the equator.  

\begin{figure}[htb]
\begin{center}
{\hspace*{-0.5cm}
\includegraphics*[height=7.0cm,width=7.5cm]{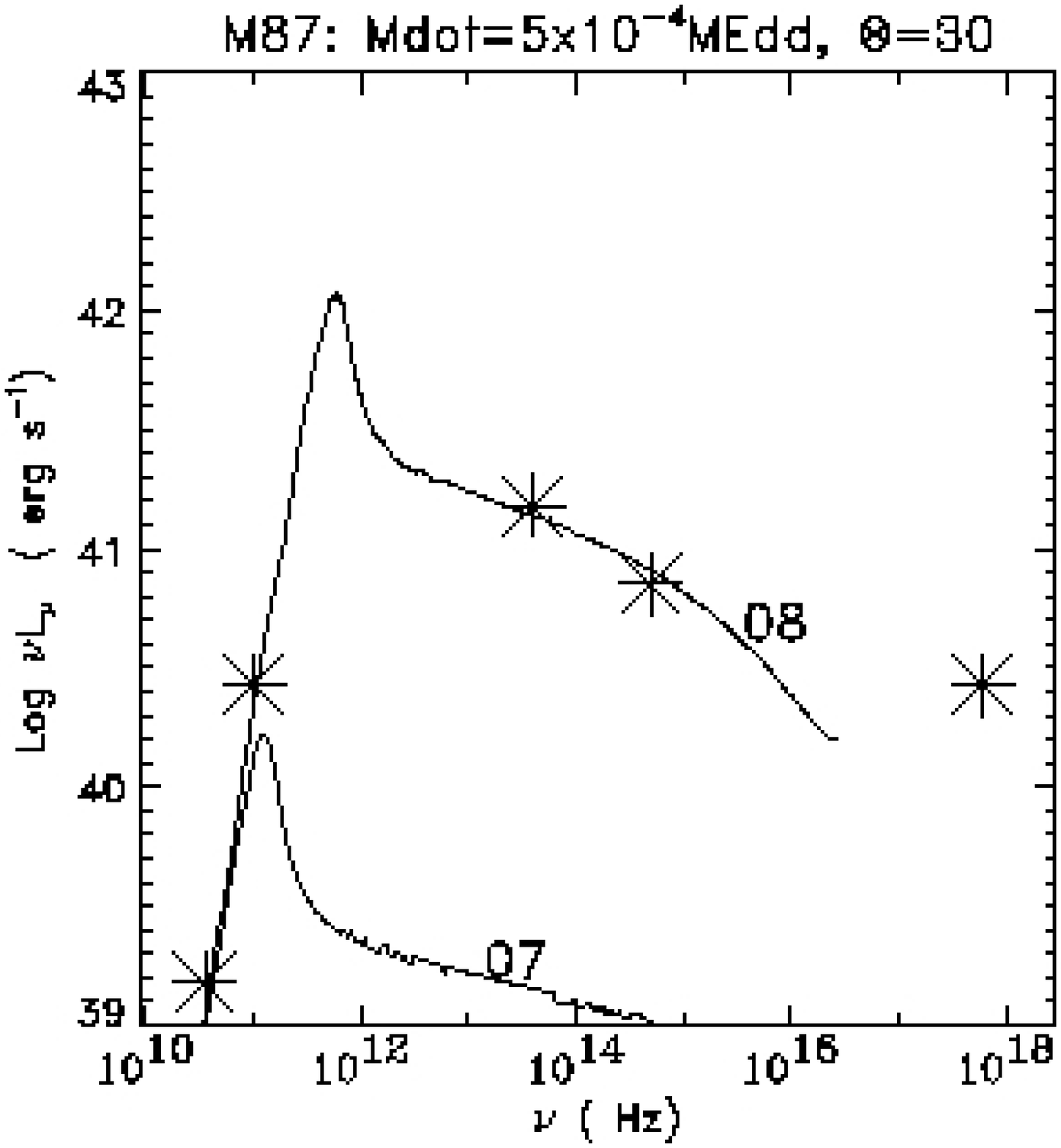}
}
\end{center}
{\vspace*{-0.4cm}}
\caption [ ] {  The SED of a disk-jet model of the elliptical galaxy M87.
   Similar to Biretta 2002 and Di Matteo 2003, an accretion rate of
   $5\times 10^{-4} \Mdot_\mathrm{Eddington}$ and $T =5\times 10^6 $K
   is set to enter the domain of calculations through the outer boundary which
   is located at 150 Schwarzschild radii from the central SMBH. The vertical
   scale height of the disk at the outer boundary is taken to be 
   $H\approx 0.1 R_\mathrm{out}$. Additionally, a hot tenuous
   corona is set to sandwich the optically thin disk.
   The calculated profiles (solid lines) are superposed on the observational
   data (asterisks). The line 07 corresponds to a model in which  the PMF
   is set to be in equipartition with the thermal energy, whereas  TMF=0.
   The line 08 is similar to 07, but the TMF is
   allowed to develop and reach values beyond equipartition with respect to the
   thermal energy of the electron in the TL.
   The above SED has been obtained by solving the radiative transfer in 4-dimensions,
   taking into account the Kompaneets equation for consistently modelling Comptonization
   (see \cite{HujCB},\cite{Huj03} for further details).   }
\end{figure}
On the other hand, the corona above the TL has been found to be dynamically unstable.
Unlike normal stars that heat up the surrounding corona from below, in the absence
of other energy sources, black holes cannot supply the surrounding corona with heat,
hence they start to collapse dynamically.
  
To elaborate this point, let us compare the conduction time scale with the dynamical
time scale along $\rm B_\mathrm{P}$-field at the last stable orbit of a 
SMBH:
\[
\DD{\tau_{\rm cond}}{\tau_{\rm dyn}} =  \DD{r \rho U_\mathrm{r}}{\kappa_0 T^{5/2}_\mathrm{i}}
 = 4.78\times 10^{-4} \rho_{10} T^{-5/2}_\mathrm{i,10} {\cal M}_8, \]
where $\rho_{10}$, $\rm{T_\mathrm{i,10}}$ and ${\cal M}_8$ 
 are respectively in $10^{-10}\,\rm{g\,cm^{-3}}$, 
 $10^{10}\,$K and in $10^8\,\Msun$ units.
This is much less than unity for most reasonable values of density and temperature typical for 
AGN-environments. In writing the above equation we have optimistically taken the upper limit $ c/\sqrt{3}$ for the velocity,
and set $\kappa_0 = 3.2\times 10^{-8}$ for the ion-conduction coefficient.
When modifying the conduction operator to respect causality, we obtain 
${\tau_{\rm cond}}/{\tau_{\rm dyn}} \le U_\mathrm{r}/c $, which is again smaller than unity.
 \\
This agrees with our numerical calculations which rule out the 
possibility of outflows  from the corona, and in particular not 
from the highly unstable polar region of the BH, where matter is neither magnetic- nor
centrifugal-supported against the central gravity of the BH.
\section{Formation of electron-proton jets: summary}

In this paper we have presented a model for initiating jets from systems
containing BHs surrounded by accretion disks. The model is based on the
following sub-structures:
\ben
\item  A weakly magnetized accretion disk in the outer region.
       Balbus-Hawley instability amplifies the
       weak MFs up to thermal equipartition which, in combination with the Parker
       instability and inward-advection of matter, re-generate a PMF of large scale topology.
       This large scale PMF turns the disk at $r_\mm{tr}$ into advection-dominated. 
       However, since this occurs on the dynamical time scale which is radius-dependent,
       $r_\mm{tr}$ is basically the outer boundary, or alternatively is close to the
       last stable orbit. The former possibility is observationally inconsistent,
       as this would force the disk to truncate at large radii and terminate accretion. 
       The later possibility is plausible: assume that 
       the flow at $r_\mm{tr}$ is freely falling, and has the vertical width of $H_d/r=1/30$.
        Using  $\rho(r,\theta=0) = \rho(r_0,\theta=0) (r_\mm{BL}/r)^{3/2}$ and 
        $\Bp(r,\theta=0) = \Bp(r_0,\theta=0) (r_\mm{BL}/r)^{2},$ 
        one can easily verify that MFs would terminate accretion if
        $r_\mm{tr} \ge 16\, r_\mm{LSO},$ 
        where $ r_\mm{LSO}$ is the last stable radius. Here we have assumed that whatever
        is the strength of the MF, the magnetized inflow interior to $r_\mm{LSO}$ 
        is gravitationally bound
        and will be accreted, together with the MF,  into the hole. 
        In this region, we anticipate large scale MFs to change into a monopole like-configuration. 
        Thus, an advective disk threaded by large scale MFs must be located inside
        $10-20\,r_\mm{LSO}$, depending on the accretion rate. 
\item  In the region  $r_\mm{top} \le r \le r_\mm{tr}$, TAWs are the dominant angular momentum
       carrier. The transport proceeds vertically and on the dynamical time scale, which 
       requires the disk, again, to be advection-dominated for stably supplying the
       jet with angular momentum. The angular momentum associated with TAWs is deposited
       in the turbulent-diffusive TL between the disk and the corona, where vast of the 
       shear-generated TMFs reconnect and heat the ions up to the virial temperature.
       Besides the fact that the large magnetic-diffusivity in the TL  
       lengthens the time scale of TAWs to cross the TL compared to the local dynamical
        time scale (which is necessary for maintaining a geometrical thin TL), 
       reconnection of the TMF-lines terminates magnetic braking and yields a 
       trapping of the angular
       momentum. Consequently, the virial-heated ions are then forced to rotate super-Keplerian
       and centrifugal-accelerate outwards.\\
       Another argument in favor of the thin geometrical structure of the TL is that the 
       corona is thermally unstable. 
       The reasons are: 1) absence of heating from below, 2)
       centrifugal forces are weak, and 3) Lorentz forces exerted on the plasma 
       in the polar region of the BH are negligibly small. 

\item The plasma in the TL is tenuous, highly advective, two-temperature and ion-dominated,
      which gives rise to the formation of electron-proton jets.
      The shear-generated TMFs in the TL are in thermal equipartition with the ions
      and in super-equipartition with the thermal energy of the electrons.
      Such a strong TMF in the TL is necessary for explaining the observed excess of 
      radio luminosities such as in M87 and GRO 1655-40.

\item  Since the pressure-gradient vanishes at the horizon, 
       a rarifaction wave starts to move from inside-to-outside, leaving  the matter
       behind its front free-falling. The resulting density profile has a global
       maximum at $r_\mm{BL} (<< r_\mm{tr}),$ and a minimum at the inner boundary  $r_\mm{in}$,
       where the dynamical time scale is shortest, and  where the strength of Coulomb coupling
       between the ions and electrons
       is weakest. Consequently, the disk truncates at  $r_\mm{BL}$ and 
       an highly advective ion-dominated torus emerges in the BL.  
       The MF in the ion-torus change topologically from large scale into monopole.
       Such a change is associated with a significant loss of magnetic flux in the hole,
      which is  necessary  for preventing magnetic termination of accretion.  
\item When does the outflow start to collimate into jet?

A significant part of the shear-generated TMFs reconnect in the TL
 and heat the ion preferentially. The other part
 is advected outwards with the plasma on the dynamical time scale.  
Inspection of Equation 1 shows that the ratio of the viscous to the advective
time scale of angular momentum in the TL reads:
\beq
\DD{\tau_\mm{visc}}{\tau_\mm{adv}} \approx \DD{r V_\mm{A}}{\nu_\mm{tur}}.
\eeq
In writing this approximation we have inserted
$V^\mm{T}_\mm{A} = B_\mm{T}/\sqrt{\rho} = V_{\varphi}$, which is applicable for the
plasma in the TL. Therefore, given the critical radius 
$r_\mm{col} (\doteq  \nu_\mm{tur}/V_\mm{A})$ at which both viscous and advective time scale
are equal, collimation will start for $r \ge r_\mm{col} $. 
This is easy to fulfil, as $V_\mm{A}$ increases with distance from the BH, or remain constant,
whereas $\nu_\mm{tur}$ must decrease. This gives rise to extremely large ratio of 
${\tau_\mm{visc}}/{\tau_\mm{adv}}$. Therefore, beyond $r_\mm{col}$, 
particle-trajectories of motion are dictated by MFs.
\een  

Finally, the model is similar to the Blandford \& Znajeck (1977) scenario, as vertical transport of angular momentum 
        by TAWs is mediated almost without advection of matter. This gives rise to high collimated
        and extremely low mass-loaded jets.  The global configuration of the jet-disk
 connection presented in this paper enhances the role of the central engine in powering the 
jet \cite{Koide}.
  A spinning BH will likely feed the jet with extra rotational energy through the frame
  dragging effect, whereas a non-rotating hole may have the opposite effect:
   extraction of energy from the plasma in the TL, hence a less powerful jet.  

Additionally,    the model  may provide explanations for several 
       other important issues related to the formation of jets and disks:
 1) the K$\alpha$ emission lines observed in 
Cyg X-1 which suggest that the disk-plasma
    in the vicinity of the last stable orbit is cold and rotates Keplerian,
      2) The origin of the super-Keplerian motions observed in the Galactic black holes
       XTE J1650-500 and MCG-6-30-15 \cite{Miller}, which
       suggest that the central objects are spinning BHs, and 
      3) the origin of the extremely low mass-loaded jets observed in several 
        AGNs and $\mu$-quasars.  
   

\def\Discussion{
\setlength{\parskip}{0.3cm}\setlength{\parindent}{0.0cm}
     \bigskip\bigskip      {\Large {\bf Discussion}} \bigskip}
\def\speaker#1{{\bf #1:}\ }
\def\endDiscussion{}


\begin{thebibliography}{99}

\bibitem{Abramowicz}  Abramowicz, M., Igumenshchev, I., Lasota, J-P.,MNRAS, 293, 433, (1998)
\bibitem{Balbus}   Balbus, S., Hawley, J.,  ApJ, {\bf 376}, (1991)
\bibitem{Biretta}  Biretta, J., Junor, W., Livio, M., 
                     New Ast. Reviews, {\bf 46}, 239, (2002)
\bibitem{Blandford1} Blandford, R., Znajek, R., MNRAS, 179, 433, (1977) 
\bibitem{Blandford2} Blandford, R., Payne, D., MNRAS, 199, 883, (1982)
\bibitem{Blandford3} Blandford, R., Begelman, M., MNRAS, 303, L1, (1999)
\bibitem{Blandford4} Blandford, R., astro-ph/0110394,  (2001)
\bibitem{Camenzind} Camenzind, M., RvMA, 3, 234, (1990)
\bibitem{Camenzind2} Camenzind, M., "The Black hole environments", Les Houches Lecture 2002,
 "Accretion Disks, Jets and high energy Phenomena in Astrophysics", eds. F. Menard et. al.
   GDPS, in press
\bibitem{DiMatteo} Di Matteo, T., Allen, S., Fabian, A., et al., ApJ, 582, 133, (2003)
\bibitem{Esin}  Esin, A., McClintock, J.E., Drake, J., et al., ApJ, 555, 483, (2001)
\bibitem{Hawley}  Hawley, F., Gammie, C., Balbus, S., ApJ, {\bf 464}, 690,  (1996)
\bibitem{Hawley}  Hawley, F., Balbus, S., ApJ, {\bf 573}, 738,  (2002)
\bibitem{Heyvearts} Heyvaerts, J, Norman, C., ApJ, 347, 1055, (1989)
\bibitem{HujCa}  Hujeirat, A., Camenzind, M., A\&A, {\bf 361}, L53,  (2000a)
\bibitem{HujCb}  Hujeirat, A., Camenzind, M., A\&A, {\bf 362}, L4,  (2000b)
\bibitem{HujR}   Hujeirat, A., Rannacher, R., New Astro. Reviews, 45, 425 (2001)
\bibitem{HujCB}  Hujeirat, A., Camenzind, M., Burkert, A., A\&A, 386, 757,   (2002)
\bibitem{HujCL}  Hujeirat, A., Camenzind, M., Livio, M., A\&A, {\bf 394}, L9,  (2002)  
\bibitem{Huj03}  Hujeirat, A., in preparation,  (2003) 
\bibitem{Ghosh}  Ghosh, P., Lamb, F.K., ApJ, 223, 83, (1978)
\bibitem{Lovelace} Lovelace, R., Berk, H., Contopoulos, J., ApJ. 379, 696, (1991)
\bibitem{Koenigl} K\"onigl, A., ApJ, 342, 208, (1989)
\bibitem{Koide} Koide, S., Shibata, K., Kudok, T., Meier, D.L., Science, 295, 1688, (2002) 
\bibitem{Miller} Miller, J.M., Fabian, A.C., Wijnands, R., ApJ, 570, L69, (2002)  
\bibitem{Mirabel00}  Mirabel, I.F., Astroph/0005591, (2000)   
\bibitem{Mirabel01}  Mirabel, I.F., ApSSS, {\bf 276}, 153, (2001) 
\bibitem{Narayan}  Narayan, R., Yi, I., ApJ, {\bf 444}, 231, (1995) 
\bibitem{Ogilvie}  Ogilvie, I., Livio, M., ApJ, {\bf 553}, 158, (2001) 
\bibitem{Popham} Popham, R., Sunyaev, R.A., ApJ, 547, 355, (2001) 
\bibitem{Pudritz}   Pudritz, R., Norman, C., ApJ, 274, 677, (1983)
\bibitem{Paczyanski} Paczyanski, B., Wiite, P.J., A\&A, 88, 23, (1980)
\bibitem{Rees}  Rees, M., Begelman, M., Blandford, R., \& Phinney, E.,  
                     Nature, {\bf 295}, 17, (1982)
\bibitem{Rybicki}  Rybicki, G., Lightman, A.P., Radiative processes, Wiley-Interscience Publication,
                  (1979)
\bibitem{Shakura}  Shakura, N. I., Sunyaev, R. A., A\&A, {\bf 24}, 337, (1973)
\bibitem{Sandnaek} Sandb{\ae}k, O., Leer, E., ApJ, 423, 500, (1994)
\bibitem{Shu88}  Shu, F., Lizano, S., Ruden, S., Najita, J., ApJ, 328, L19, (1988)
\bibitem{Shu91} Shu, F., The Physics of Star Formation and Early stellar Evolution, ed. Kylafis, N.D.,
                \& Lada, C.J., Dordrecht: Kluwer), 365, (1991) 
\bibitem{Shu94}  Shu, F., Najita, J., Ruden, S., Lizano, S., ApJ, 429, 781, (1994)
\bibitem{Simon}  Simon, T., ASP, 223, 235, (1999)
 \bibitem{Spiegel} Spiegel, E., Zahn, J.-P., A\&A, 265, 106, (1992)
\bibitem{Stone}  Stone, J.M.,  Pringle, J.E., Astrph/0009233, (2000)
\bibitem{Tout}  Tout, C.A., Pringle, J.E., MNRAS, 259, 604, (1992)
\bibitem{Uchida} Uchida, Y., Shibata, K., PASJ, 37, 515, (1985)
\bibitem{Yuan} Yuan, F., Markoff, S., Falke, H., 383, 854, (2002)
\end{thebibliography}
\end{document}